\documentclass[reprint,superscriptaddress,pre]{revtex4-1}

\usepackage{todonotes}
\usepackage{subfig}
\usepackage{graphicx}
\usepackage{amsmath,amssymb}
\usepackage{units}
\usepackage[version=3]{mhchem}
\usepackage[percent]{overpic}

\newcommand\drawaxes{ \put(-5,-5){\vector(1,0){20}} \put(-5,-5){\vector(0,1){20}} \put(-6.5,17){a} \put(17,-6.5){c}}

\begin{document}

\title{Coherency strain and the kinetics of phase separation in \ce{LiFePO4}}
\author{Daniel A. Cogswell}
\affiliation{Department of Chemical Engineering, Massachusetts Institute of Technology, Cambridge, Massachusetts 02139, USA}
\author{Martin Z. Bazant}
\affiliation{Department of Chemical Engineering, Massachusetts Institute of Technology, Cambridge, Massachusetts 02139, USA}
\affiliation{Department of Mathematics, Massachusetts Institute of Technology, Cambridge, Massachusetts 02139, USA}
\date{\today}

\begin{abstract}
A theoretical investigation of the effects of elastic coherency on the thermodynamics, kinetics, and morphology of intercalation in single \ce{LiFePO4} nanoparticles yields new insights into this important battery material.  Anisotropic elastic stiffness and misfit strains lead to the unexpected prediction that low-energy phase boundaries occur along \{101\} planes, while conflicting reports of phase boundary orientations are resolved by a partial loss of coherency in the \{100\} direction. Elastic relaxation near surfaces leads to the formation of a striped morphology, whose characteristic length scale is predicted by the model and yields an estimate of the interfacial energy. The effects of coherency strain on solubility and galvanostatic discharge are studied with a reaction-limited phase-field model, which quantitatively captures the influence of misfit strain, particle size, and temperature on solubility seen in experiments. Coherency strain strongly suppresses phase separation during discharge, which enhances rate capability and extends cycle life.  The effects of elevated temperature and the feasibility of nucleation are considered in the context of multi-particle cathodes.
\end{abstract}

\maketitle

Lithium iron phosphate (\ce{LiFePO4}) has emerged as an important high-rate cathode material for rechargeable batteries \cite{Tang2010} and is unique because of its strongly anisotropic diffusivity \cite{Morgan2004,Islam2005}, its strong elastic anisotropy \cite{Maxisch2006}, and its tendency to phase-separate into lithium-rich and lithium-poor phases \cite{Padhi1997,Delacourt2005,Dodd2006,Yamada2006,Meethong2007}.  Despite a few conclusive observations of phase boundaries in \textit{chemically} delithiated \ce{LiFePO4} nanoparticles \cite{Chen2006,Laffont2006,Ramana2009}, the general consensus has been that phase boundaries always form during \textit{electrochemical} discharge, thereby limiting battery performance \cite{Padhi1997,Srinivasan2004}.  However, this limitation is inconsistent with dramatic rate improvements resulting from smaller nanoparticles, doping \cite{Chung2002}, and surface coatings \cite{Kang2009}.

The feasibility of phase boundary formation has recently been challenged by both phase-field models \cite{Kao2010,Bai2011} and ab-initio calculations \cite{Malik2011}.  In a companion paper \cite{Bai2011}, we demonstrate that high discharge currents can suppress phase separation in reaction-limited nanoparticles, so that the spinodal is a \textit{dynamic} property of intercalation systems.  In this paper we consider the additional effect of elastic coherency strain and find that it leads to a quantitatively accurate phase-field description of \ce{Li_xFePO4} that is useful for interpreting experimental data.  Via mathematical analysis and numerical simulations of galvanostatic discharge, we conclude that coherency strain strongly suppresses phase separation, leading to better battery performance and improved mechanical durability.

Coherency strain arises when molar volume is a function of composition, i.e. due to the difference in lattice parameters between \ce{FePO4} and \ce{LiFePO4}.  Two-phase systems with identical crystal structure and small misfit strains generally form coherent interfaces \cite{Cahn1961,Cahn1962}.  It has been suggested that \ce{Li_xFePO4} retains coherency throughout nucleation and growth \cite{Meethong2007}, and in-situ observation of crystalline material during battery operation supports this prediction \cite{Delmas2008,Leriche2010}.

As we show, the observation of aligned phase boundaries and striped morphologies in \ce{Li_xFePO4} \cite{Chen2006,Laffont2006,Ramana2009} provide conclusive evidence of coherency strain.  Furthermore, it is necessary to consider the fully anisotropic elastic constants and misfit strain to interpret experiments.  Simplified elastic analysis has led to the conclusion that \{100\} is always the preferred orientation \cite{Meethong2007a,VanderVen2009}, although \{101\} phase boundaries are sometimes observed \cite{Laffont2006,Ramana2009}.  Our fully anisotropic analysis predicts that \{101\} is the low-energy orientation, and we attribute the observation of \{100\} boundaries to a partial loss of coherency resulting from dislocations (or cracks).

The origin of striped morphologies \cite{Chen2006} has not been satisfactorily explained. It has been suggested that they result from the characteristic wavelength of spinodal decomposition \cite{Ramana2009}, although it is not clear why the instability would be frozen in this state.  We show instead that stripes represent the stable equilibrium state of finite size particles, and predict that the spacing scales with the square root of particle size. As a result, we are able to extract the interfacial energy from experimental micrographs.

The reported solubilities of Li in \ce{FePO4} vary significantly and depend on particle size and temperature \cite{Yamada2005,Chen2006,Yamada2006,Meethong2007,Meethong2007a,Delmas2008,Kobayashi2009,Badi2011,Wagemaker2011}.  These differences in solubility can now be explained in light of coherency strain. Phase field calculations of solubility as a function of particle size and temperature are able to fit experimental data with just two parameters.

An alternative to coherent phase separation is for entire particles to remain homogeneous and form a mosaic pattern, with some particles existing at low concentration and others at high concentration \cite{Dreyer2010,Dreyer2011,Delmas2008}.  This scenario is energetically favorable since there is no phase boundary energy or change in solubility, but it requires exchange of material between nanoparticles.  By constructing phase diagrams for both the coherent and mosaic scenarios, we find a limited role for coherent nucleation and growth and predict that moderately elevated temperatures ought to suppress all two-phase coexistence, even in large particles.

\subsection*{Phase-field model}
We begin with the reaction-limited phase-field model for ion-intercalation in single, anisotropic nanoparticles from our prior work~\cite{Singh2008,Burch2008,Bai2011} and extend it to include coherency strain.  The theory couples electrochemical surface reactions to bulk phase separation using a thermodynamically consistent generalization of Butler-Volmer kinetics. The reaction rate depends on the Cahn-Hilliard  \cite{Cahn1958} (or Van der Waals~\cite{vdw1893}) gradient energy, introduced to model the formation of phase boundaries.

Individual nanoparticles are modeled as open thermodynamic systems in contact with an electrolyte mass reservoir at constant temperature, volume, and chemical potential (grand canonical ensemble).  For solids, PV work is generally very small and can be neglected at atmospheric pressure.  Thus we consider the free expansion of the particle at zero applied pressure.  The inhomogeneous grand free energy functional is \cite{Cahn1961,Cahn1962}:
\begin{equation}
  \Xi[c(\vec{x}),\vec{u}(\vec{x})]=\int_V f(c)-\mu c+\frac{1}{2}\kappa(\nabla c)^2+\frac{1}{2}C_{ijkl}\epsilon_{ij}\epsilon_{kl}\, dV
 \label{Eq:energy_functional}
\end{equation}
where $c(\vec{x})$ is the mole fraction of lithium, $f(c)$ is the homogeneous Helmholtz free energy density, $\mu c$ is a Legendre tranform that accounts for the chemical potential $\mu$ of lithium ions in the reservoir, $\kappa$ is the gradient energy coefficient that introduces interfacial energy, and $\frac{1}{2}C_{ijkl}\epsilon_{ij}\epsilon_{kl}$ is elastic strain energy.  $C_{ijkl}$ is the elastic stiffness tensor, and $\epsilon_{ij}(\vec{x})$ is the total strain field which may be decomposed into three parts:
\begin{equation}
 \epsilon_{ij}(\vec{x})=\bar{\epsilon}_{ij}+\frac{1}{2}\left(u_{i,j}+u_{j,i}\right)-\epsilon_{ij}^0c(\vec{x})
\end{equation}
where $\bar{\epsilon}_{ij}$ is a homogeneous strain that accounts for uniform deformation of the particle, $\frac{1}{2}\left(u_{i,j}+u_{j,i}\right)$ is a local inhomogeneous correction to $\bar{\epsilon}_{ij}$ resulting from compositional inhomogeneity, and $\epsilon_{ij}^0$ is the lattice misfit between \ce{FePO4} and \ce{LiFePO4} (a stress-free strain).  We assume that the misfit strain varies linearly with composition (Vegard's Law).

The electrochemical reaction at the surface of the particle is governed by a generalized Butler-Volmer equation  for solids and concentrated solutions~\cite{Bazant_notes,Burch2009,Bai2011}:
\begin{align}
  J&=J_0\left[e^{-\alpha\frac{e\eta}{kT}}-e^{(1-\alpha)\frac{e\eta}{kT}}\right] &  J_0&=k_0\frac{a_+^{1-\alpha}a^\alpha}{\gamma_A}
\end{align}
where the overpotential has a variational definition:
\begin{equation}
 \eta(c,\nabla^2 c)=\Delta\phi - \Delta\phi_{eq}=\frac{\delta\Xi}{\delta c(\vec{x})}
\end{equation}
$\Delta\phi_{eq}$ is the Nernst equilibrium potential and $\Delta\phi=\phi_e-\phi=\frac{\mu}{e}$ is the interfacial voltage, where $\phi$ and $\phi_e$ are the electrostatic potentials of ions and electrons, respectively.  When a potential $\Delta\phi$ is applied, the system is displaced from equilibrium and lithium enters or leaves the system.  

It is a subtle but important point that we define overpotential relative to the Nernst voltage rather than the voltage plateau of the phase-separated system at zero current.  The Nernst voltage is an equilibrium material property, but the voltage plateau is not.  A flat voltage plateau is commonly cited as a hallmark of phase separation, but as we will demonstrate, coherency strain leads to upward-sloping plateaus. 

To model the experimentally relevant case of galvanostatic discharge, the current flow into the particle is constrained by an integral over the active area at the surface of the particle:
\begin{align}
 I&=\int_A \frac{\partial c}{\partial t}\, dA &  \frac{\partial c}{\partial t}&=J_0\left[e^{-\alpha\frac{e\eta}{kT}}-e^{(1-\alpha)\frac{e\eta}{kT}}\right]+\xi
 \label{Eq:evolution}
\end{align}
where $\xi$ is a Langevin noise term.  Mechanical equilibrium equations must additionally be solved for the displacement vector $\vec{u}(\vec{x})$:
\begin{align}
 \nabla\cdot\sigma_{ij}&=0 &  \int_V C_{ijkl}\epsilon_{ij}(\vec{x})\, dV&=0
 \label{Eq:mechanical_equilibrium}
\end{align}
The second equation defines the average stress in the system to be zero, which is necessary for stress-free boundaries and equivalent to minimizing $\Xi$ with respect to $\bar{\epsilon}_{ij}$.

\begin{figure*}[t]
 \centering
  \subfloat[]{\includegraphics[width=.3\textwidth]{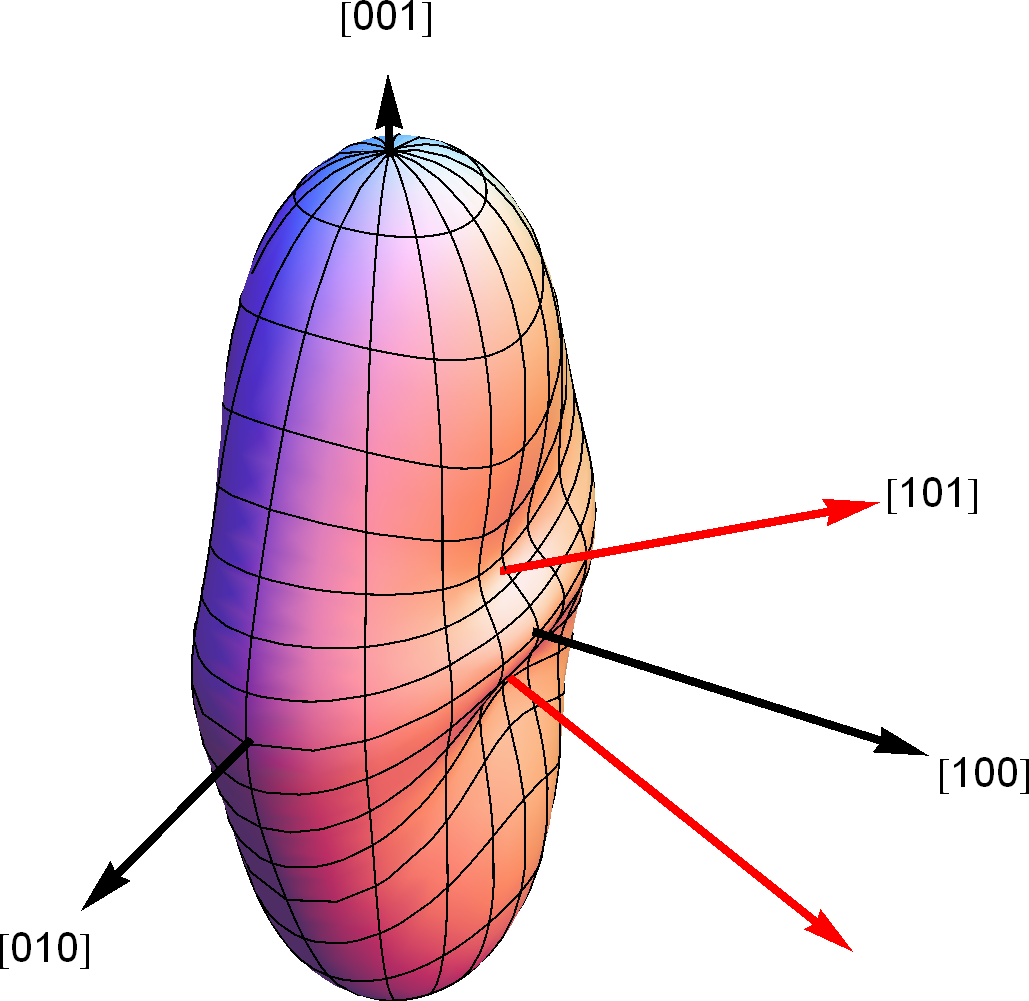}
   \label{Fig:B_a}}
  \subfloat[]{\includegraphics[width=.3\textwidth]{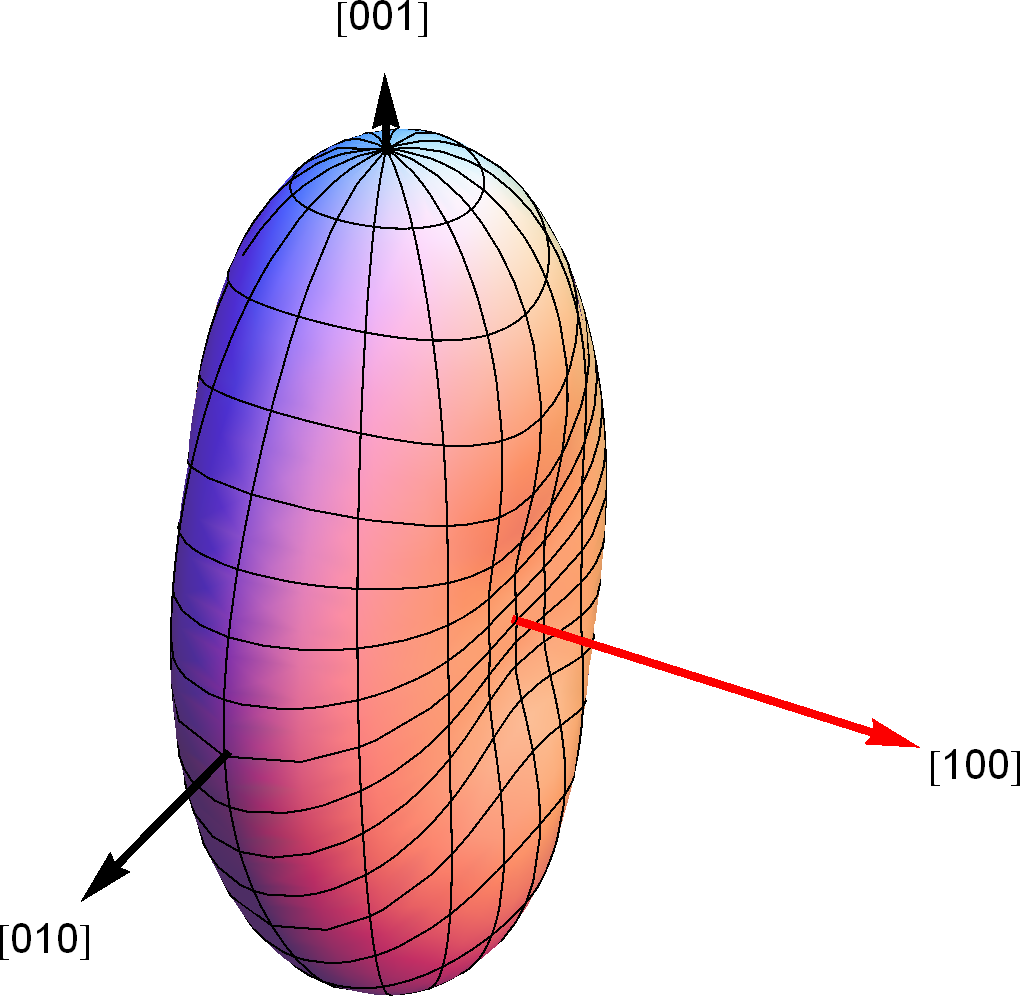}
   \label{Fig:B_b}}
  \subfloat[]{\includegraphics[width=.3\textwidth]{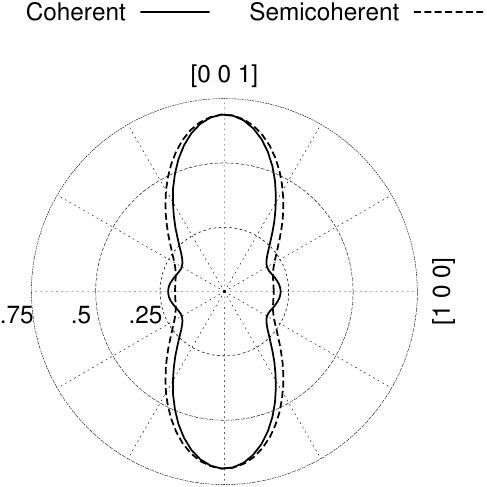}
   \label{Fig:B_c}}
 \caption{Spherical plots of $B(\vec{n})$, the elastic strain energy of a flat interface as a function of normal direction for \subref{Fig:B_a} a coherent interface and \subref{Fig:B_b} a semicoherent interface which has lost coherency in the [001] direction.  The red arrows indicate $\vec{n}_0$,  the direction of minimum energy.
  \subref{Fig:B_c} A polar plot of $B(\vec{n})$ in the in the a-c plane comparing coherent and semicoherent energies.  Energy is in units of GPa.}
 \label{Fig:B}
\end{figure*}

Small particles are expected to be limited by surface reactions, and in this case Eq. \ref{Eq:evolution} and \ref{Eq:mechanical_equilibrium} constitute a depth-averaged system of equations that can be solved in 2D on the particle's active surface.  The validity of the depth-averaged approximation is supported by anisotropic elastic considerations (Fig. \ref{Fig:B}) and by phase-field simulation of diffusion in the lithium channels \cite{Tang2011}.

\section*{Results}
\subsection*{Phase boundary orientation}
Since \ce{LiFePO4} is orthorhombic, it is necessary to consider its fully anisotropic elastic stiffness and the anisotropic lattice mismatch between phases when analyzing phase boundary morphology.  Assuming that the elastic modulus of each phase is the same (homogeneous modulus assumption), Khachaturyan \cite{Khachaturyan1967,Khachaturyan2008} related the elastic energy of an arbitrarily anisotropic elastic inclusion to a function of direction: 
\begin{equation}
 B(\vec{n})=C_{ijkl}\epsilon_{ij}^0\epsilon_{kl}^0-\vec{n}_i\sigma_{ij}^0\Omega_{jl}(\vec{n})\sigma_{lm}^0\vec{n}_m
 \label{Eq:B}
\end{equation}
$\vec{n}$ is the interface normal, and $\Omega$, which is related to the elastic Green's tensor, is defined by its inverse tensor $\Omega_{ij}^{-1}=C_{iklj}\vec{n}_k\vec{n}_l$.  Elastic energy is a function of orientation $\vec{n}$ because a phase boundary produces zero strain in the normal direction.  The direction $\vec{n}_0$ that minimizes Eq. \ref{Eq:B} defines the habit plane, which is the elastically preferred orientation of the phase boundary that minimizes strain energy.

Equation \ref{Eq:B} is plotted in Fig. \ref{Fig:B_a} using anisotropic elastic constants calculated via first-principles \cite{Maxisch2006} and anisotropic lattice mismatch measured in \cite{Chen2006}.  The figure reveals that \{010\} and \{001\} interfaces are high-energy, which justifies {\it a posteriori} the depth-averaged approximation in our model~\cite{Singh2008}, based on fast diffusion and no phase separation in the \{010\} depth direction.  The orientation of $\vec{n}_0$ obtained by numerical minimization is drawn in red.  There are four minima which lie along the \{101\} family of crystal planes, where $B(\vec{n}_0)=\unit[.19]{GPa}$.  Fig. \ref{Fig:stripes} compares the predicted \{101\} phase boundaries with experimental observations by Ramana et al.  \cite{Ramana2009}.  Laffont et al. also appear to have observed a \{101\} interface in Fig. 2b of \cite{Laffont2006}.

On the other hand, \{100\} interfaces have  been reported in some experiments \cite{Chen2006,Ramana2009}, which according to Fig. \ref{Fig:B_a}, should not be elastically preferred.  The resolution to this apparent discrepancy may be the formation of dislocations, which lead to a loss of coherency in the [001] direction.  Indeed, Chen et al.~\cite{Chen2006} report observing cracks and dislocations running in the [001] direction of \textit{negative} misfit strain.

Stanton and Bazant \cite{Stanton2011} recognized the importance of negative misfit for \ce{LiFePO_4} with isotropic elastic analysis, but here we consider the fully anisotropic case.  Fig. \ref{Fig:B_b} plots $B(\vec{n})$ for a semicoherent interface with $\epsilon_{33}=0$, and Fig. \ref{Fig:B_c} compares the coherent and semicoherent cases in cross section.  The semicoherent habit plane lies along the \{100\} family of planes, and curiously $B(\vec{n}_0)$ remains nearly unchanged by to the loss of coherency.  The orientation of the interface changes, but its elastic energy does not.

The mechanism by which coherency is lost remains to be determined.  It could be that the phases initially form coherently, but then lose coherency over time lose as dislocations form.  It is also possible that phase boundaries form semicoherently upon lithiation,  aligned with pre-existing cracks or defects.  The dynamics of phase separation for the latter scenario is presented in Fig. \ref{Fig:phase_separation}.

\subsection*{Morphology}
\begin{figure}[t]
 \subfloat[500x500\,nm]{\drawaxes \includegraphics[width=.45\columnwidth]{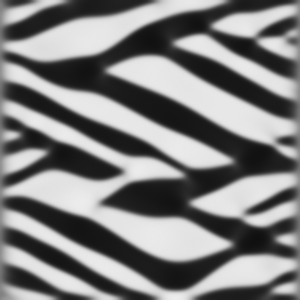}
  \label{Fig:stripes_a}}
 \hspace{.02\columnwidth}
 \subfloat[500x500\,nm]{\drawaxes \includegraphics[width=.45\columnwidth]{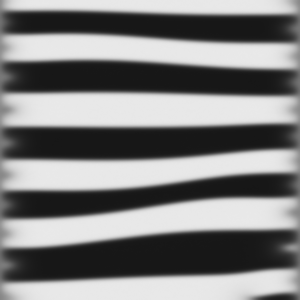}
  \label{Fig:stripes_b}}\\
 \subfloat[]{\includegraphics[width=.45\columnwidth]{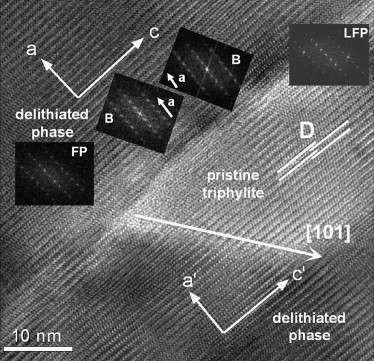}
  \label{Fig:stripes_c}}
 \hspace{.02\columnwidth}
 \subfloat[]{\includegraphics[width=.45\columnwidth]{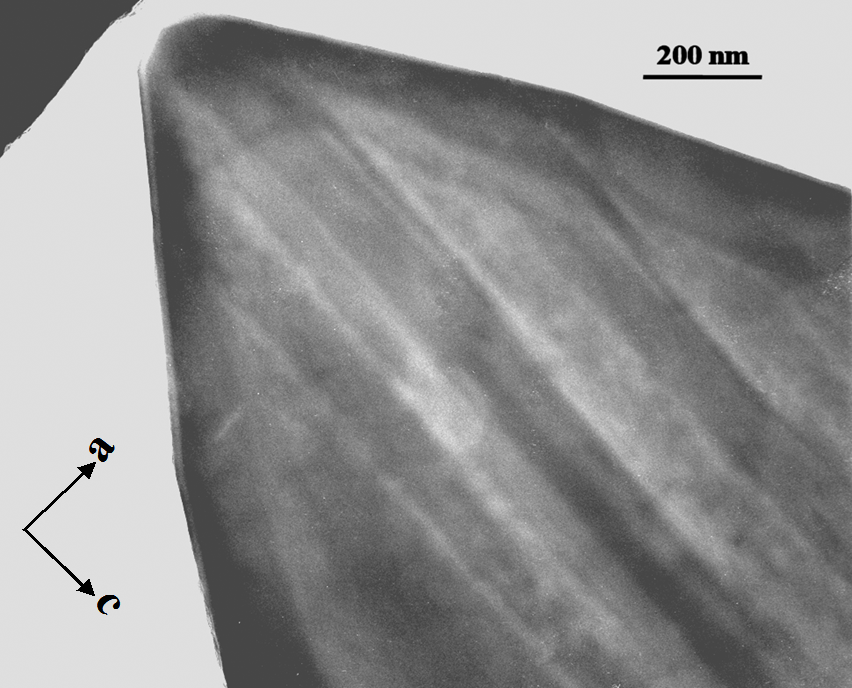}
  \label{Fig:stripes_d}}
 \caption{Comparison of simulated and experimental microstructures in \ce{Li_{.5}FePO4}.  See Fig. \ref{Fig:phase_separation} for simulation dynamics.
  \subref{Fig:stripes_a} Phase boundaries align along \{101\} planes to minimize elastic coherency strain.
  \subref{Fig:stripes_b} Loss of coherency in the [001] direction causes the stripes form along \{100\} planes.
  \subref{Fig:stripes_c} HRTEM image of a \{101\} phase boundary.  Reprinted from \cite{Ramana2009} with permission from Elsevier.
  \subref{Fig:stripes_d} TEM image of \{100\} stripes \cite{Chen2006}.  Reproduced by permission of The Electrochemical Society.}
 \label{Fig:stripes}
\end{figure}

Modulated structures resulting from coherency strain are often observed in experimental systems, and stripes are an equilibrium morphology that minimizes energy in finite size particles \cite{Khachaturyan1969,Khachaturyan2008}.  Stripes form due to elastic relaxation at the surface of the particle, and align normal to $\vec{n}_0$.  The characteristic wavelength $\lambda$ of the stripes balances the elastic energy of surface relaxation which scales with volume, and total interfacial energy which scales with particle size \cite{Khachaturyan1969,Khachaturyan2008}.  Evidence of this relaxation is visible near the boundaries of the simulated particle in Fig. \ref{Fig:stripes_b}.

The wavelength of periodicity is described by a scaling relation derived in the supporting material:
\begin{equation}
  \lambda=\sqrt{\frac{2\gamma L_c}{\Delta f}}
  \label{Eq:scaling}
\end{equation}
$\lambda$ is the period of the striping, $\gamma$ is interfacial energy, $L_c$ is the width of the particle in the [001] direction, and  $\Delta f$ is the difference in free energy density between the homogeneous state and the coherent phase-separated state.  $\Delta f$ has a chemical contribution from the homogeneous free energy density $f(c)$, and an elastic contribution from coherency strain (Eq. \ref{Eq:strain_energy}). We find  $\Delta f=\unit[4.77]{MJ/m^3}$ using the regular solution model and gradient energy that were fitted to experimental data in Methods.

Phase-field simulation and experimental observation of stripes are compared in Fig. \ref{Fig:stripes_a} and \ref{Fig:stripes_c}.  Using Eq. \ref{Eq:scaling}, the striping in Fig \ref{Fig:stripes_d} can be used to obtain the \ce{FePO4}/\ce{LiFePO4} interfacial energy.  Applying Eq. \ref{Eq:scaling} to the striped pattern in Fig. \ref{Fig:stripes_d}, with $L_c=\unit[4]{\mu m}$ and $\lambda\approx\unit[250]{nm}$  (measured away from the corner to mitigate the influence of particle geometry), we infer a phase boundary energy of $\unit[37]{mJ/m^2}$.  For the phase-field simulation in Fig. \ref{Fig:stripes_b}, $L_c=\unit[500]{nm}$, $\lambda\approx \unit[90]{nm}$, and $\gamma=\unit[39]{mJ/m^2}$.  The interfacial energy can also be calculated directly from the phase-field model \cite{Cahn1958,Cahn1962a}, which yields $\gamma=\unit[39]{mJ/m^2}$, in agreement with the rule of thumb that coherent phase boundaries have interfacial energies less than $\unit[200]{mJ/m^2}$ \cite{Porter2009}.  This confirms both the validity of the scaling relation and our choice of phase-field parameters (in particular $\kappa$, which has until now been difficult to estimate).

Fig. \ref{Fig:phase_separation} shows the dynamics of phase separation for a homogeneous particle that is held at zero current.  Both the coherent and semicoherent cases are shown.  The initial decomposition is followed by a period of coarsening, but coarsening stops when the stripes reach their characteristic wavelength, which scales with $\sqrt{L_c}$.  Thus the stripes are dependent on particle geometry and do not coarsen, as would be expected if they were related to the must unstable wavelength of spinodal decomposition\cite{Ramana2009}.

\subsection*{Critical particle size}
Phase-field methods have been used in studies of size-dependent solubility without coherency strain \cite{Nauman1989,Burch2009}, and a minimum system size was found below which two-phase coexistence is prohibited.  This minimum size is set by the diffuse width of the phase boundary.  Here we find that the critical particle size criterion changes with the introduction of coherency strain, and is a result of the combined effects of coherency strain and interfacial width.

\begin{figure}[t]
 \centering
 \subfloat[]{
  \begin{overpic}[width=.4\textwidth]{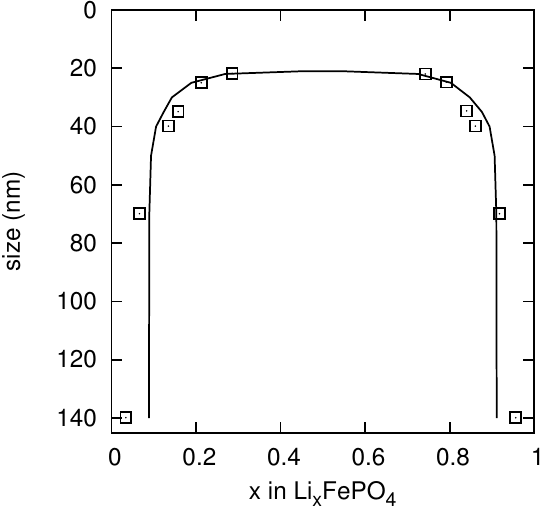}
   \put(47,50){\includegraphics[width=.1\textwidth]{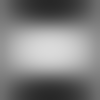}}
   \put(49,45){30x30\,nm}
  \end{overpic}
  \label{Fig:Wagemaker_miscibility}}\\
 \subfloat[]{\includegraphics[width=.4\textwidth]{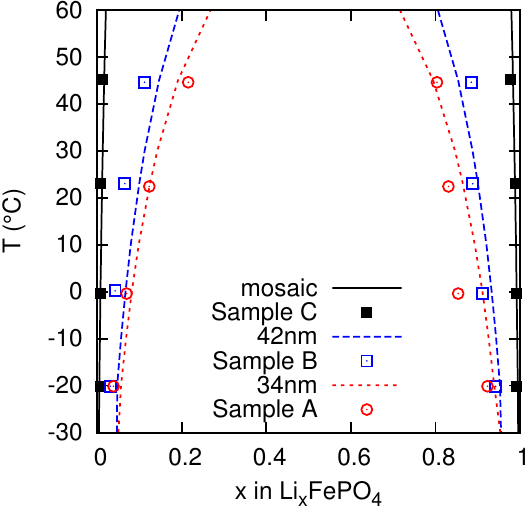}
  \label{Fig:Meethong_miscibility}}
 \caption{Comparison of calculated and measured \ce{LiFePO4} solubility limits as a function of temperature and particle size.
  \subref{Fig:Wagemaker_miscibility} Size dependence at room temperature.  Data points from \cite{Wagemaker2011}.
  \subref{Fig:Meethong_miscibility} Size and temperature dependence.  Data points from \cite{Meethong2007}.}
 \label{Fig:miscibility}
\end{figure}

Figure \ref{Fig:miscibility} compares phase-field calculations and measurements of the solubility limits as a function of particle size and temperature.  A regular solution was used for $f(c)$, and the regular solution parameter $\Omega$ and gradient energy $\kappa$ were obtained with a least-squares regression of the phase-field model to the data points in Fig. \ref{Fig:miscibility} (see Methods).  With just these two parameters we were able to simultaneously fit both size and temperature dependence of lithium solubility (four experimental data sets).  This confirms that the \ce{Li$_x$FePO$_4$} system may reasonably be described as a regular solution.

The fitting in Fig. \ref{Fig:Meethong_miscibility} offers new insight into the experiments themselves.  In Ref. \cite{Meethong2007}, the differences in the miscibility gap were originally thought to be related to particle size, but according to  Fig. \ref{Fig:miscibility}, \unit[100]{nm} and \unit[42]{nm} particles should not show significantly different solubilities.  This assertion is confirmed by x-ray diffraction (XRD) analysis in the paper, which found strain in samples A and B, but not in C.  Sample A and B were prepared differently than C, which may explain why phase boundaries did not form in the sample C particles (see below).  Fig. \ref{Fig:Meethong_miscibility} shows that the shrinking miscibility gap in samples A and B is plausibly explained by coherency strain.

The equilibrium phase boundary width was measured from simulation to be \unit[12]{nm}, which is in good agreement with the 12-\unit[15]{nm} width measured by STEM/EELS \cite{Laffont2006}.  In phase-field simulations of small particles near the critical size, we observed that phase separation always occurs as a sandwich (see Fig. \ref{Fig:Wagemaker_miscibility} inset), sometimes with the lithiated phase in the middle, and other times with the delithiated phase in the middle.  Presumably this is a result of elastic interaction between the phases.  Both cases require the formation of two interfaces, explaining why the critical particle size of \unit[22]{nm} is roughly twice the interfacial thickness.

\subsection*{Phase diagram}
\label{Sec:phase_diagram}
\begin{figure*}[t]
 \centering
  \includegraphics[width=.8\textwidth]{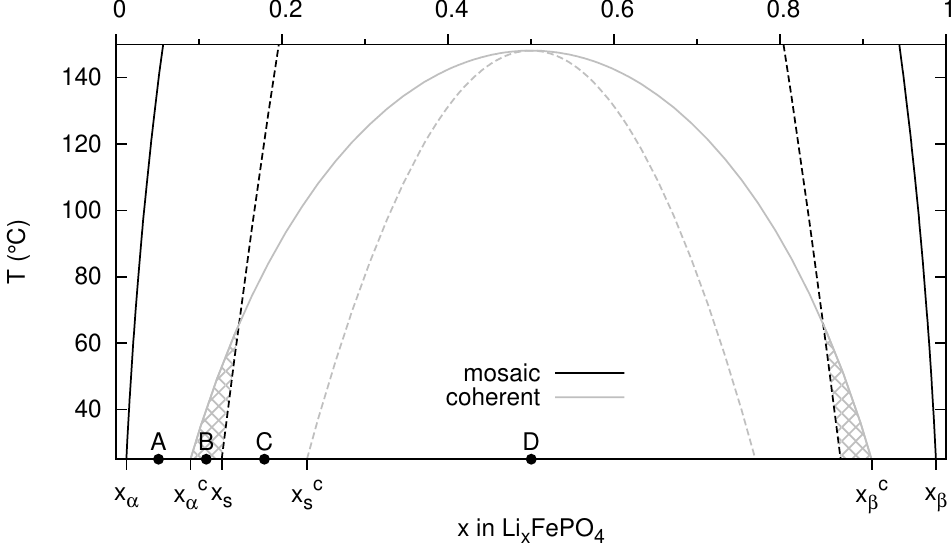}\\
  \hspace{.05\textwidth}\includegraphics[width=.8\textwidth]{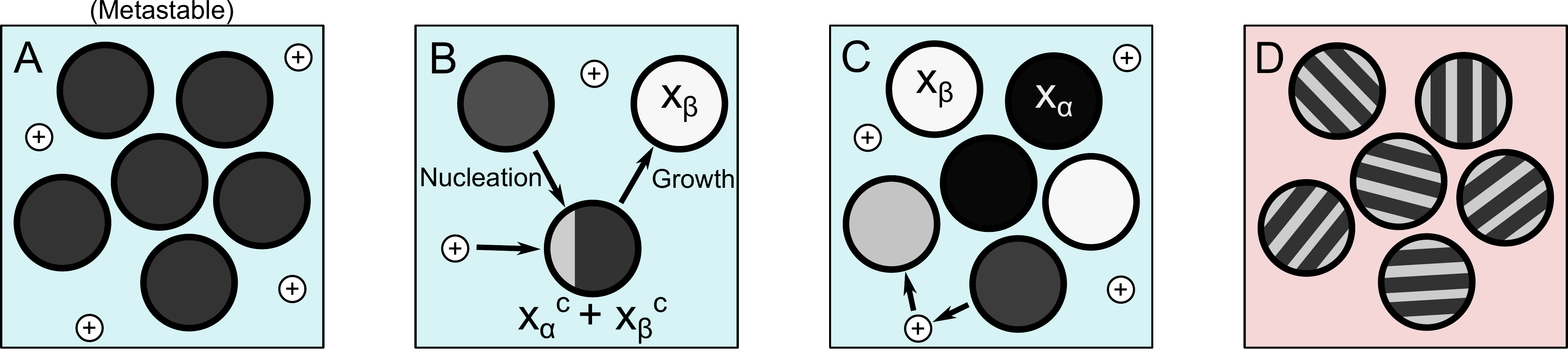}
 \caption{Phase diagram comparing coherent and mosaic phase separation.   Solid lines are the limits of miscibility, and dashed lines indicate the spinodal.  Hashing denotes the region where coherent nucleation is feasible.  The illustrations depict a fully delithiated cathode that has been discharged to points A,B,C and D in the phase diagram.  A, B, and C are bathed in electroyte (light blue), but D is not.}
 \label{Fig:phase_diagram}
\end{figure*}

Fig. \ref{Fig:phase_diagram} shows a phase diagram that was calculated using the fitted regular solution model (See Methods).  A mosaic phase diagram was calculated using $f(c)$, and a coherent phase diagram was calculated by adding elastic energy ( Eq. \ref{Eq:strain_energy}) to $f(c)$.  The eutectoid reaction \cite{Delacourt2005,Dodd2006} involving a disordered phase at higher temperatures  has been neglected.  The phase diagram reveals that coherency strain stabilizes the solid solution at temperatures above $150^{\circ}$C, well below the disordering temperature.

The illustrations in Fig. \ref{Fig:phase_diagram} depict a completely delithiated cathode nanoparticle that has been discharged to the corresponding points in the phase diagram.  At point A the particles are inside the mosaic miscibility gap, but do not transform since there is no phase transformation pathway.  The microstructure is thus metastable with respect to mosaic decomposition.  At point B, the particles cross the coherent spinodal and coherent nucleation inside particles becomes possible.  Phase transformation will proceed slowly in this region since nucleation is an activated process, and for fast discharge this region will be bypassed.

By point C the particles have crossed the mosaic spinodal, and spontaneously form a mosaic if they are able to exchange ions through the electrolyte (light blue).  Current plays as important role in the onset of this transition, since the position of the spinodal moves inward and eventually disappears with increasing current \cite{Bai2011}.

Point D illustrates the case where discharge occurs rapidly to a point inside the coherent spinodal without time for exchange of ions between particles.  The red background color indicates that the particles are not in electrolyte, and therefore cannot exchange ions.  The particles relax to a phase-separated state in this case only.  Chemical delithiation, which has produced observations of phase boundaries, corresponds to case D.

The difference in solubilities predicted in  Fig. \ref{Fig:phase_diagram} can be used as a guide to the interpretation of experimental data.  If a mosaic forms, the existence of a second phase first becomes possible inside the mosaic solubility limits, and particles will have compositions of either $x_\alpha=.01$ or $x_\beta=.99$.  XRD measurements of a system of fully intercalated and fully deintercalated individual particles confirms the appearance of a second phase by $x=.04$ at room temperature \cite{Delmas2008}, and equilibrium measurements find very little room temperature solubility \cite{Yamada2005,Meethong2007}.

However if coherent phase boundaries form inside particles, the onset of phase separation in Fig. \ref{Fig:phase_diagram} will only occur for $.09<x<.91$.  Frequent reports of extended regions of solid solution \cite{Chen2006,Yamada2006,Meethong2007,Meethong2007a,Kobayashi2009,Badi2011,Wagemaker2011} support this prediction, and several authors \cite{Meethong2007a,Kobayashi2009} have attributed their observations to retained strain.  Moreover, the XRD measurements of Chen et al. \cite{Chen2006} on striped particles did not detect phase-separation until at least 10\% of the lithium had been extracted.   Badi et al. \cite{Badi2011} also recently measured the composition of coexisting Li$_\alpha$FePO$_4$ and  Li$_\beta$FePO$_4$  phases, and found $\alpha\approx .1$ and  $\beta\approx .9$.  Both of these observations agree precisely with the predicted coherent miscibility gap in Fig. \ref{Fig:phase_diagram} at room temperature.

\subsection*{Phase separation dynamics}
Spinodal decomposition in systems with coherency strain was studied by Cahn \cite{Cahn1961,Cahn1962}, who found that for homogeneous systems at equilibrium, strain energy can be approximated for small fluctuations as:
\begin{equation}
 \frac{1}{2}C_{ijkl}\epsilon_{ij}\epsilon_{kl}\approx \frac{1}{2}B(\vec{n}_0)(c-x)^2
 \label{Eq:strain_energy}
\end{equation}
Since strain energy is a function of the mean composition of the system, coherency strain invalidates the common tangent construction and leads to an upward-sloping voltage plateau \cite{VanderVen2009}.

A linear stability analysis of the evolution equations (Eq. \ref{Eq:evolution}) was performed in \cite{Bai2011}, and the amplification factor was found to be: 
\begin{equation}
s=-\sqrt{\bar{J}_0+\frac{I^2}{4}}(\bar{\mu}'+\kappa k^2)+I\left(\frac{\bar{J}_0'}{\bar{J}_0}+\frac{1}{2}\kappa k^2\right)
\end{equation}
Bar notation indicates evaluation of functions at the homogeneous state $c=x$.  The coherent diffusion potential is found by inserting Eq. \ref{Eq:strain_energy} into Eq. \ref{Eq:energy_functional} and taking the variational derivative:
\begin{equation}
 \mu=f'(c)-\kappa\nabla^2c+B(\vec{n}_0)(c-x)
\end{equation}

\begin{figure}[t]
 \subfloat[]{\includegraphics[width=.8\columnwidth]{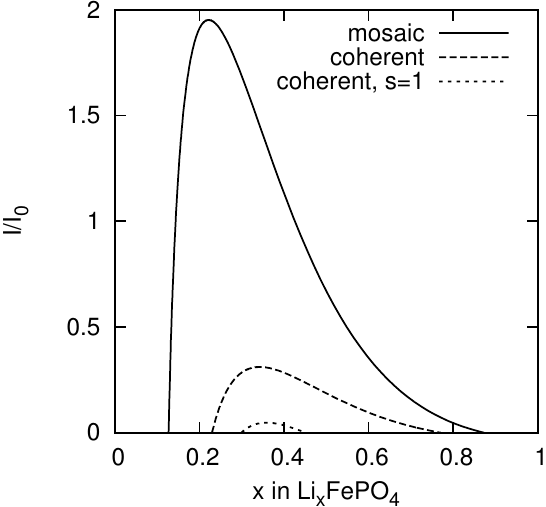}
  \label{Fig:discharge_stability}}\\
 \subfloat[]{\includegraphics[width=.8\columnwidth]{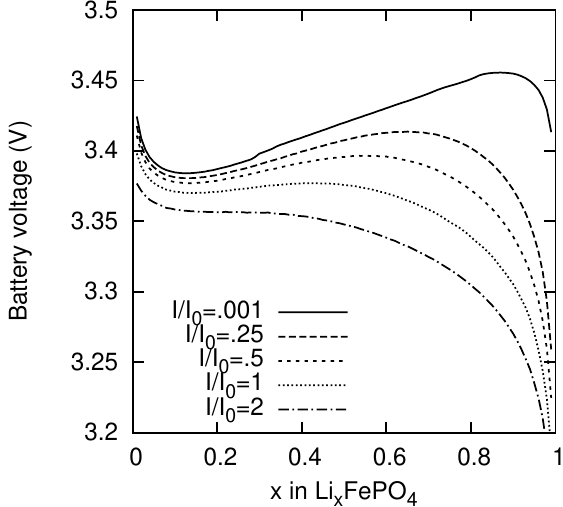} \label{Fig:discharge_curves}}\\
 \subfloat[$\frac{I}{I_0}=.001$]{\includegraphics[width=.22\columnwidth]{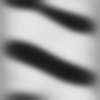} \label{Fig:discharge_b}}
 \hspace{.01\columnwidth}
 \subfloat[$\frac{I}{I_0}=.01$]{\includegraphics[width=.22\columnwidth]{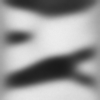} \label{Fig:discharge_c}}
 \hspace{.01\columnwidth}
 \subfloat[$\frac{I}{I_0}=.033$]{\includegraphics[width=.22\columnwidth]{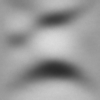} \label{Fig:discharge_e}}
 \hspace{.01\columnwidth}
 \subfloat[$\frac{I}{I_0}=.05$]{\includegraphics[width=.22\columnwidth]{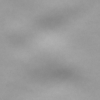} \label{Fig:discharge_d}}
 \caption{Analysis and simulation of galvanostatic discharge.
  \subref{Fig:discharge_stability} Linear stability boundary as a function of current.
  \subref{Fig:discharge_curves} Constant-current discharge curves for a single nanoparticle with coherent interfaces.
  \subref{Fig:discharge_b}-\subref{Fig:discharge_e}  Phase boundary morphology showing formation of a quasi-solid-solution as discharge rate is increased  (\ce{Li_{.6}FePO4}, 100x100\,nm).}
 \label{Fig:discharge}
\end{figure}

Linear stability with and without coherency strain is presented in Fig. \ref{Fig:discharge_stability}, and shows a transition from phase-separating to homogeneous filling as current increases.  Coherency strain promotes the stability of the solid solution by reducing the critical current and shrinking the spinodal.  The neutral stability curves  in the figure represent the boundary between stable and unstable dynamics ($s=0$),  and the $s=1$ curve is where the amplification factor is large enough to produce phase separation on the order of the discharge time.  This curve is an indicator of when complete phase separation is observable, and has a maximum at $\frac{I}{I_0}=.047$.  In between the coherent curve and the $s=1$ curve is a region of {\it quasi-solid-solution} where there are unstable modes, but not enough time for complete phase separation~\cite{bai2011}.

The transition from fully phase-separating to quasi-solid-solution is captured in the simulated microstructures of Fig. \ref{Fig:discharge} at $x=0.6$ for filling at different currents.  By $\frac{I}{I_0}=.05$ (slightly above the maximum of the $s=1$ curve), phase separation is just barely visible.  Thus we conclude that due to coherency strain, phase separation is suppressed when the applied current exceeds only a few percent of the exchange current.

Voltage curves during discharge were calculated at different currents and are presented in Fig. \ref{Fig:discharge}.  The most striking difference compared to the incoherent case  \cite{Bai2011} is the upward-sloping voltage plateau when phase separation occurs, which was predicted at equilibrium by Van der Ven et al. \cite{VanderVen2009}.  The first ions to enter the particle do extra mechanical work straining the surrounding lattice, and this work is recovered by the last ions, which enter lattice sites that have already been partially strained.  When phase separation is suppressed at higher currents however, there are no phase boundaries and hence coherency strain does not play any role.

\begin{figure*}[t]
 \centering
 \subfloat[]{\drawaxes \includegraphics[width=.22\linewidth]{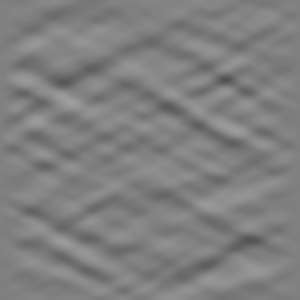}
  \label{Fig:phase_separation_a}}
 \hspace{.02\linewidth}
 \subfloat[]{\includegraphics[width=.22\linewidth]{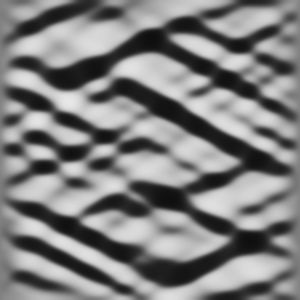}}
 \hspace{.02\linewidth}
 \subfloat[]{\includegraphics[width=.22\linewidth]{c000100.png}}
 \hspace{.02\linewidth}
 \subfloat[]{\includegraphics[width=.22\linewidth]{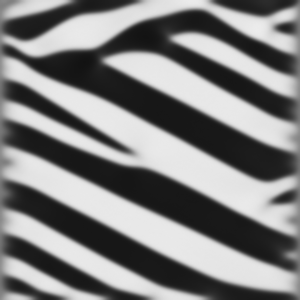}
  \label{Fig:phase_separation_d}}\\
 \subfloat[]{\drawaxes \includegraphics[width=.22\linewidth]{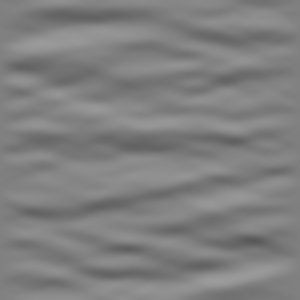}
  \label{Fig:phase_separation_e}}
 \hspace{.02\linewidth}
 \subfloat[]{\includegraphics[width=.22\linewidth]{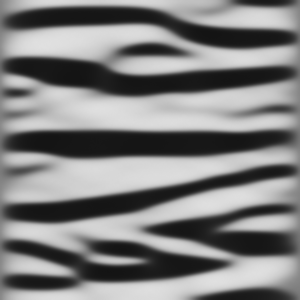}}
 \hspace{.02\linewidth}
 \subfloat[]{\includegraphics[width=.22\linewidth]{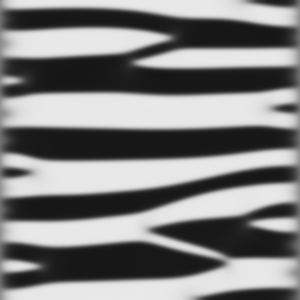}}
 \hspace{.02\linewidth}
 \subfloat[]{\includegraphics[width=.22\linewidth]{semi_c000400.png}
  \label{Fig:phase_separation_h}}
 \caption{
  An initially homogeneous system (\ce{Li_{.5}FePO4}, 500x500\,nm) decomposes into regions of high and low lithium concentration when held at zero current.
 \subref{Fig:phase_separation_a}-\subref{Fig:phase_separation_d} The phase boundaries align along \{101\} planes to minimize elastic coherency strain.
  \subref{Fig:phase_separation_e}-\subref{Fig:phase_separation_h} Loss of coherency in the [001] direction results produces \{100\} interfaces.}
 \label{Fig:phase_separation}
\end{figure*}

\section*{Discussion}

In this paper we have presented a thermodynamically consistent phase-field model for nanoparticulate intercalation materials and focused on the significance of coherency strain in the $\ce{Li_xFePO4}$ two-phase system.  With just two free parameters (the regular solution parameter $\Omega$ and the gradient energy $\kappa$), our model simultaneously explains the observed phase boundary orientations, stripe morphologies, the measured phase boundary width, interfacial energy, size- and temperature-dependent solubilities, and reports of extended solid solution.  Elastic analysis reveals that \{101\} is the preferred phase boundary, that negative misfit strain along the [001] axis explains the observation of \{100\} phase boundaries, and that elastic relaxation at the surface of the particles is the origin of stripes.  Analysis and simulation of galvanostatic discharge shows that coherency strain significantly suppresses phase separation during discharge and leads to upward-sloping voltage curves for single particles below the critical current.

Although phase boundaries have been observed experimentally, we conclude that most electrochemical data is inconsistent with phase boundaries forming inside \ce{LiFePO4} nanoparticles.  The method of sample preparation appears to strongly influence the feasibility of phase boundary formation.  We suspect that chemical delithiation may be partly responsible for the observation of phase boundaries by preventing the exchange of lithium between particles.  Chemical delithiation \cite{Yamada2006, Chen2006,Ramana2009,Kobayashi2009,Wagemaker2011}, a high degree of Li antisite defects  \cite{Badi2011}, particle size, and the synthesis technique itself \cite{Meethong2007,Meethong2007a} all appear to influence the formation of phase boundaries.

According to  Fig. \ref{Fig:phase_diagram}, temperature may play an important role in improving battery performance.  Most battery research focuses on room temperature operation, but room temperature discharge passes through the hatched region of Fig. \ref{Fig:phase_diagram} where nucleation is possible and the coherent miscibility is crossed before the mosaic spinodal.  However for temperatures greater than $70^\circ C$, the situation reverses, and the mosaic instability occurs before coherent nucleation.  Therefore, moderately elevated temperatures could be useful for stabilizing the solid solution.  Homogeneous particles have advantages for battery performance since they have larger active area for insertion and do not waste energy forming phase boundaries.  Homogeneous particles also avoid the internal stresses caused by phase boundaries, which benefits cycle life.

By deriving a simple scaling relation, we have been able to estimate the phase boundary energy by measuring the stripe wavelength in an experimental micrograph.  Our mathematical formula predicts $\gamma=\unit[39]{mJ}$ in precise agreement with the calculated value from the phase-field model.  The fact that $\gamma$ is small may have significant consequences for the role of nucleation at small currents.  Cahn and Hilliard showed that nucleation and growth competes with spinodal decomposition near the limit of metastability \cite{Cahn1959,Cahn1962}.  In the case of Fig. \ref{Fig:phase_diagram}, the energy barrier for nucleation in the hashed region is important.  Applying the approximation of Cahn and Hilliard, we estimate the room temperature energy of coherent homogeneous nucleation to be $\unit[102]{kT}$ at the mosaic spinodal.  Although this energy is fairly large, heterogeneous nucleation at the particle surface is a more likely pathway in small nanoparticles with a large surface to volume ratio.  The heterogeneous nucleation barrier is likely to be only a fraction of the homogeneous barrier, placing heterogeneous nucleation well within the realm of kinetic relevance.  Energy barriers must generally be less than $\unit[78]{kT}$ for observable rates of nucleation \cite{Porter2009}.  Therefore our future work will focus on accurately calculating the critical nucleus and heterogeneous nucleation barrier energy in order to understand the role of nucleation.

\section*{Methods}
\subsection*{Numerical methods}
Eq. \ref{Eq:evolution} and \ref{Eq:mechanical_equilibrium} were solved in 2D as a coupled system of differential-algebraic equations using finite difference methods on a square grid.  The Matlab function ode15s was used for time integration with the integral constraint implemented using a singular mass matrix.  Zero-flux boundaries were applied to $c(\vec{x})$, and zero pressure boundaries were applied to $\vec{u}$.  At each timestep it was necessary to solve for $c(\vec{x}), u_i(\vec{x}), u_j(\vec{x}), \Delta\phi, \bar{\epsilon}_1,\bar{\epsilon}_2, \bar{\epsilon}_3$, and $\bar{\epsilon}_6$.

\subsection*{Phase-field parameters}
The inputs to the phase-field model that must be selected for the \ce{LiFePO4} system are the homogeneous free energy $f(c)$, the gradient energy coefficient $\kappa$, the elastic stiffness $C_{ijkl}$, and the lattice mismatch $\epsilon_{ij}^0$.  We use GGA+U first-principles calculations of $C_{ijkl}$ for \ce{FePO4} \cite{Maxisch2006}, and lattice mismatch measured in \cite{Chen2006}.  For $f(c)$ we assume regular solution model:
\begin{equation}
 V_{Li}f(c)=\Omega c(1-c)+kT(c\ln(c)+(1-c)\ln(1-c))
 \label{Eq:regular_solution}
\end{equation}
where $V_{Li}$ is the volume per lithium atom in the solid.  The regular solution parameter $\Omega$ and the gradient energy $\kappa$ were obtained with a least-squares regression of the phase-field model to experimental measurements of the miscibility gap \cite{Meethong2007, Wagemaker2011}, illustrated in Fig. \ref{Fig:miscibility}.  Phase-field simulations were performed by allowing a square particle at $x=.5$ to relax to equilibrium at zero current.  The miscibilities were then found by taking the minimum and maximum compositions in the equilibrated microstructure.

The best fit was achieved with $\Omega=\unit[115]{meV/Li}$ ($\unit[11.1]{kJ/mol}$) and $\kappa=\unit[3.13\times 10^9]{eV/m}$  ($\unit[5.02\times 10^{-10}]{J/m}$).  The root-mean-square error was $2.3\%$.  Both values are close to those used by Tang et al. \cite{Kao2010,Tang2011}.  Sample C in Fig. \ref{Fig:Meethong_miscibility} had no measured strain and large particles that do not exhibit a size effect.  Thus it is influenced by $f(c)$ only.  The fact that a particularly good fit is achieved with sample C supports the use of a regular solution for \ce{Li_xFePO4}.

\subsection*{Stripe scaling}

\begin{figure}[t]
 \centering
 \subfloat[]{\includegraphics[width=.7\columnwidth]{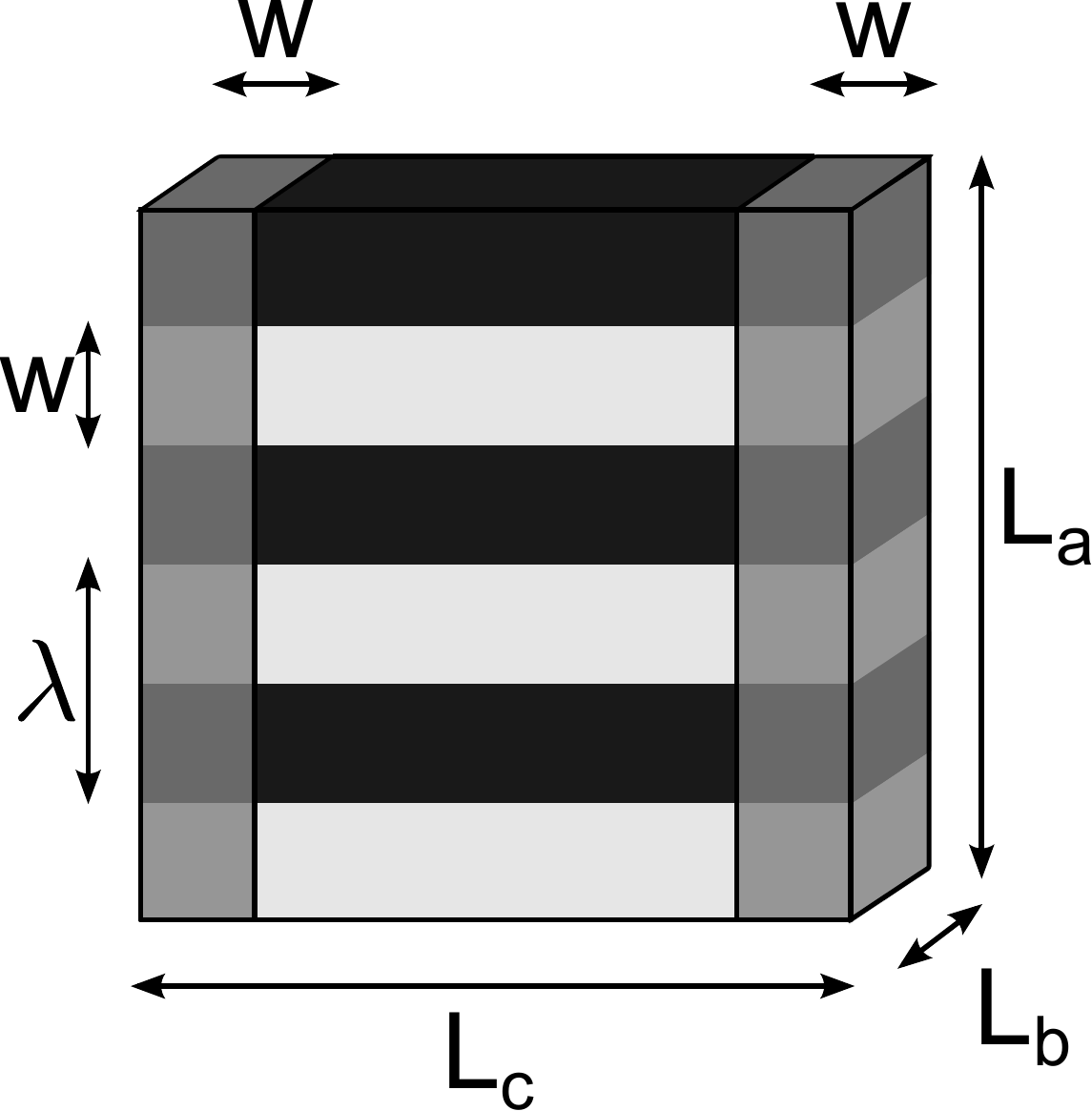}
  \label{Fig:scaling}}\\
 \subfloat[]{
  \includegraphics[width=.8\columnwidth]{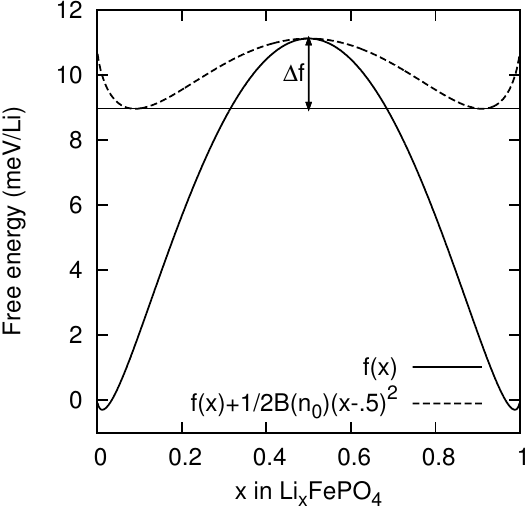}
  \label{Fig:free_energy}}
 \caption{
  \subref{Fig:scaling} Illustration of the stripe morphology in \ce{Li_{.5}FePO_4}.  Elastic relaxation occurs in the shaded regions at the \{001\} surfaces of the particle.
  \subref{Fig:free_energy} The influence of coherent phase separation on free energy.}
\end{figure}

Here we adapt the arguments of Khachaturyan \cite{Khachaturyan1969,Khachaturyan2008} to derive an expression for the period of striping in finite size \ce{Li_{.5}FePO_4} particles.  Illustrated in Fig. \ref{Fig:scaling}, the stripes form to balance elastic relaxation at the \{001\} surfaces of the particle (energy/volume) and total interfacial energy (energy/area).  The change in energy due to elastic relaxation ($\Delta f=\unit[4.77]{MJ/m^3}$) is illustrated in Fig. \ref{Fig:free_energy}, and is found with a common tangent construction applied to $f(c)+\frac{1}{2}B(\vec{n}_0)(c-x)^2$, as described in \cite{Cahn1961,Cahn1962,Cahn1962a}.  Assuming that the width of relaxation $w$ at the boundary is comparable to the size of the stripes, the change in energy due to relaxation and creation of phase boundaries is:
\begin{equation}
 \Delta E=2\Delta f L_a L_b w+\gamma\frac{L_a}{w} L_b L_c
\end{equation}
The equilibrium stripe size will minimize $\Delta E$, and so we solve for $w$ when $\frac{d\Delta E}{dw}=0$:
\begin{equation}
 \frac{d\Delta E}{dw}=2\Delta f L_a L_b-\frac{\gamma L_a L_b L_c}{w^2}=0
\end{equation}
Solving for the stripe period $\lambda=2w$, we obtain:
\begin{equation}
 \lambda=2w=\sqrt{\frac{2\gamma L_c}{\Delta f}}
\end{equation}
The $\sqrt{L_c}$ dependence reveals that striping is an effect of finite size domains.  As the domain size approaches infinity, the equilibrium state approaches two infinite domains separated by one interface.

\begin{acknowledgments}
We are grateful to Jacob White for insightful discussions on numerical methods.  This work was supported by the National Science Foundation under Contracts DMS-0842504 and DMS-0948071 and by a seed grant from the MIT Energy Initiative.
\end{acknowledgments}

\bibliography{Nature_LFP_coherency}
\end{document}